\documentclass[letterpaper, 10 pt, conference]{ieeeconf}  % Comment this line out if you need a4paper

\IEEEoverridecommandlockouts                              % This command is only needed if 
\usepackage{graphicx}
\usepackage{graphics} % for pdf, bitmapped graphics files
\usepackage{epsfig} % for postscript graphics files
\usepackage{mathptmx} % assumes new font selection scheme installed
\usepackage{times} % assumes new font selection scheme installed
\usepackage{amsmath} % assumes amsmath package installed
\usepackage{amssymb}  % assumes amsmath package installed
\usepackage{kotex}

\usepackage{hyperref}
\hypersetup{
    colorlinks=true,
    linkcolor=blue,
    filecolor=magenta,      
    urlcolor=blue,
}

\title{\LARGE \bf
Fisheye-Calib-Adapter: An Easy Tool for Fisheye Camera Model Conversion}

\author{Sangjun Lee$^{1}$% <-this % stops a space
\thanks{This work has been submitted to the IEEE for possible publication. Copyright may be transferred without notice, after which this version may no longer be accessible.}
\thanks{$^{1}$Sangjun Lee is with Visual Positioning Team, StradVision, Seoul, Korea
        {\tt\small eowjd0512@gmail.com}}%
}

\begin{document}

\maketitle
\thispagestyle{empty}
\pagestyle{empty}

%%%%%%%%%%%%%%%%%%%%%%%%%%%%%%%%%%%%%%%%%%%%%%%%%%%%%%%%%%%%%%%%%%%%%%%%%%%%%%%%
\begin{abstract}

The increasing necessity for fisheye cameras in fields such as robotics and autonomous driving has led to the proposal of various fisheye camera models. While the evolution of camera models has facilitated the development of diverse systems in the field, the lack of adaptation between different fisheye camera models means that recalibration is always necessary, which is cumbersome.
This paper introduces a conversion tool for various previously proposed fisheye camera models. It is user-friendly, simple, yet extremely fast and accurate, offering conversion capabilities for a broader range of models compared to existing tools. We have verified that models converted using our system perform correctly in applications such as SLAM. By utilizing our system, researchers can obtain output parameters directly from input parameters without the need for an image set and any recalibration processes, thus serving as a bridge across different fisheye camera models in various research fields. We provide our system as an open source tool available at: \href{https://github.com/eowjd0512/fisheye-calib-adapter}{https://github.com/eowjd0512/fisheye-calib-adapter}

\end{abstract}

%%%%%%%%%%%%%%%%%%%%%%%%%%%%%%%%%%%%%%%%%%%%%%%%%%%%%%%%%%%%%%%%%%%%%%%%%%%%%%%%


\section{INTRODUCTION}

Fisheye cameras are utilized in fields such as robotics and autonomous driving due to their wide Field of View (FoV), which provides more environmental information than pinhole cameras \cite{fisheye-application}. They are particularly used in technologies such as Visual Odometry and Simultaneous Localization And Mapping (SLAM) for estimating mobile motion (\cite{FisheyeORBSLAM, FisheyeDSO, FisheyeDVO}). In the field of computer vision, there is active research on training neural networks using fisheye images (\cite{fisheye1, fisheye2, fisheye3}).

Defining a fisheye camera model is crucial as it allows various problems to be solved mathematically by utilizing a model that accurately represents the fisheye camera. Recently, several fisheye camera models have been proposed (\cite{UCM, EUCM, DS, OCC, KB}). Each new model's introduction typically leads to new versions of systems that utilize it. Applying datasets to these systems invariably requires the camera model coefficients, thus necessitating a calibration process.

Calibration is essential to obtain a fisheye camera model. It involves acquiring the model's coefficients through the correspondence between a known object's actual size and its 2D image pixels. Calibration can be achieved using target-based methods like checkerboards or non-target-based methods leveraging landmark information such as the Manhattan assumption (\cite{fisheye-calibration, self-calibration}).

However, there are scenarios where calibration is not feasible. For instance, if a dataset provided for research only includes coefficients for a specific fisheye camera model and does not provide a calibration dataset, it is impossible to perform calibration. This scenario often occurs when a dataset is proposed that is fixed to a specific model. Moreover, when attempting experiments with this dataset, challenges arise due to the inability to perform comparisons with systems that have applied new camera models.

To address these situations, we propose a fisheye camera model adapter that allows direct conversion among the most commonly used fisheye camera models today: UCM \cite{UCM}, EUCM \cite{EUCM}, Double Sphere \cite{DS}, Kannala-Brandt \cite{KB}, OCamCalib \cite{OCC}, and Radial-Tangential distortion model \cite{RT} for pinhole camera model. Utilizing projection and unprojection, it requires only the coefficients of the model to be converted and does not necessitate any images for the recalibration process.

The contributions of our paper are as follows:

\begin{itemize}
\item We provide a simple tool for direct adaptation among a variety of fisheye camera models.
\item We detail the projection and unprojection processes between camera models, and propose methods for optimization, including initialization techniques, cost functions, and Jacobians, making it applicable to various systems.
\item We also offer an interface that facilitates the application of different models.
\end{itemize}

The paper is structured as follows: Section \ref{sec:related work} discusses Related Work, Section \ref{sec:method} describes the Method, Section \ref{sec:experiment} covers Experiments, and Section \ref{sec:conclude} concludes the paper.

% Related work
\section{Related Work}
\label{sec:related work}
\subsection{Fisheye Camera Models}

3D points are projected onto the image plane by a predefined camera model \cite{multiview-geometry}. Conversely, 2D pixel points on the image plane can be restored into 3D rays by unprojecting them using the camera model. 
The fundamental projective model, known as the pinhole model, can employ the Radial-Tangential \cite{RT} distortion model to account for distortion. However, the model is designed to suit cameras with narrow fields of view, such as pinhole cameras, making it less effective for wide-angle coverage. To address this, several models have been defined for fisheye cameras, which are an extension of the pinhole camera.

The Kannala-Brandt model \cite{KB} (widely used in OpenCV \cite{opencv} for a fisheye camera model), also considered as the equidistant distortion model for pinhole cameras \cite{kalibr}, has been proposed for wide-angle fisheye lens distortion. It focuses on a polynomial model of radial distortion, omitting the tangential distortion term. Similarly, the OCamCalib model \cite{OCC} includes an affine transformation term to correct sensor misalignment.

Models such as Radial-Tangential, Kannala-Brandt, and OCamCalib require the estimation of coefficients for polynomial terms, thus necessitating many parameters. The Unified Camera Model (UCM) \cite{UCM} serves as a catadioptric camera model capable of modeling pinhole and fisheye cameras with a single distortion parameter using parabolic, hyperbolic, elliptic, and planar mirrors. However, to perfectly model fisheye cameras, additional distortion parameters were needed, leading to the proposal of the Enhanced Unified Camera Model (EUCM) \cite{EUCM} and the Double Sphere cite{DS} model.

\subsection{Fisheye Camera Model Conversion}

We have recently discovered research related to fisheye camera model conversion \cite{libPer}. This system proposes conversions among three models: Kannala-Brandt, UCM, and OCamCalib. However, conversions are only possible through a dependency relationship from Kannala-Brandt to UCM, and from UCM to OCamCalib. Our proposed method enables direct conversions among the Kannala-Brandt, UCM, EUCM, Double Sphere, OCamCalib, and RT models without any dependencies.

% Method
\section{Method}
\label{sec:method}

\begin{figure}[!tb]
  \centering
  \includegraphics[width=3in]{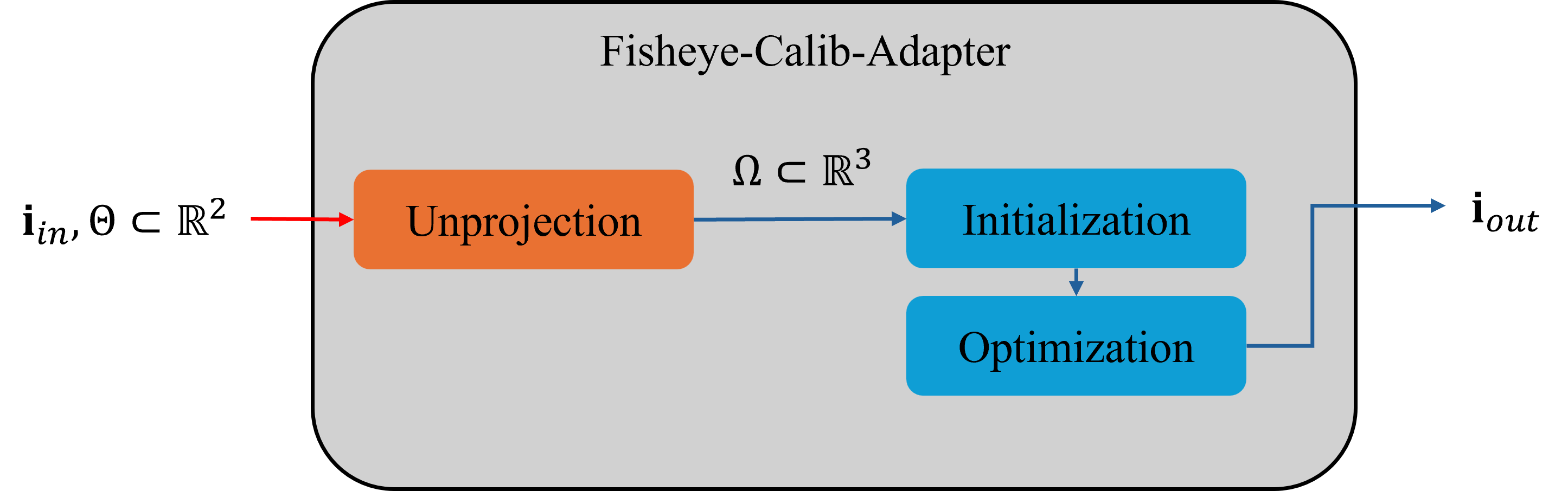}

  \caption{System overview}
  \label{figure1}
\end{figure}

The proposed Fisheye Camera Model Adapter (FCA) undergoes a process as illustrated in Figure \ref{figure1}.
As shown in Figure \ref{figure1}, the FCA receives an input model and exports an output model. The transformation is represented as:
\begin{equation}
\textbf{i}_{out} = \text{FCA}(\textbf{i}_{in})
\end{equation}
where $\textbf{i}_{in}$ and $\textbf{i}_{out}$ represent the parameters of the input and output models, respectively.

The FCA module unprojects $N$ sampled points based on the given input model and uses these derived 3D points to perform initialization and optimization for the output model. Both the initialization and optimization processes are driven by the projection function of the output model.

This modeling is possible under the assumption that the input camera model and the target output camera model were used to capture images in the same environment. The recovered ray from the input model will project to the same location regardless of where along the ray the depth is placed. Furthermore, since the projected point is used for recovering in the input model, its pair information is already known. Utilizing this, arbitrary points on the ray can be matched with their projected counterparts, allowing for the estimation of the output camera model.

In our proposed FCA, the camera models handled for conversion are defined as follows. We have defined the most commonly used fisheye camera models recently, which include:
Kannala-Brandt, 
Unified Camera Model, 
Enhanced Unified Camera Model, 
Double Sphere, 
OCamCalib, 
and the Radial-Tangential distortion model for pinhole cameras. Additionally, we address cases of other variant models in the Custom model section.

The subsequent sections of the Method introduce the unprojection function and the initialization and optimization methods for each camera model.

\noindent $\textbf{Projection function}$ Given intrinsic parameters and coefficients $\mathbf{i}$, the projection function is defined as $\mathbf{\pi}(\mathbf{x},\mathbf{i}): \Omega \xrightarrow{} \Theta$, where $\mathbf{x}=[x,y,z]^T \in \Omega\subset \mathbb{R}^3$.

\noindent $\textbf{Unprojection function}$  The unprojection function converts image coordinates back to a unit-length bearing vector as $\mathbf{\pi}^{-1}(\mathbf{u}, \mathbf{i}): \Theta \xrightarrow{} \mathbb{S}^2$, which defines a ray onto which all points corresponding to these image coordinates are projected, where $\mathbf{u}=[u, v]^T \in \Theta \subset \mathbb{R}^2$.

% \begin{figure}[thpb]
%       \centering
%       \includegraphics[width=2in]{figure5.png}

%       %\includegraphics[scale=1.0]{figurefile}
%       \caption{Inductance of oscillation winding on amorphous
%        magnetic core versus DC bias magnetic field}
%       \label{figurelabel}
%    \end{figure}
   
\noindent $\textbf{Initialization}$
For the parameters $\chi$ that we wish to optimize, we solve the linear equation $\mathbf{A}\chi=\mathbf{b}$, where $\mathbf{A}$ and $\mathbf{b}$ are derived from the projection function.

\noindent $\textbf{Optimization}$  The parameters of the output model are obtained by solving the following nonlinear least squares:  

\begin{align}
\mathbf{i}^* &= \text{arg}\min_{\textbf{i}} \sum_n^N \textbf{e}(\textbf{u}_n, \textbf{i})^T \Lambda \textbf{e}(\textbf{u}_n, \textbf{i}),
\end{align}
where the residual is
\begin{align}
\mathbf{e}(\textbf{u}, \mathbf{i}_{out}) &= \mathbf{\pi}(\mathbf{\pi}^{-1}(\mathbf{u}, \mathbf{i}_{in}),\mathbf{i}_{out})-\mathbf{u}.
\end{align}

Since the $\mathbf{u}$ corresponding pairs used for computing the input and output models are identical, the information matrix $\Lambda$ utilizes the identity matrix. Additionally, given $\mathbf{u}$, the condition satisfying the projection and unprojection for $\langle\mathbf{u}, \tilde{\mathbf{x}} \rangle$, where $\tilde{\mathbf{x}} = \mathbf{\pi}^{-1}(\mathbf{u}, \mathbf{i}_{in})$,  is always provided, allowing the error term for the optimization to be defined as:

\begin{equation}
\label{generale}
\mathbf{e}(\mathbf{i}_{out})=\mathbf{\pi}(\tilde{\mathbf{x}},\mathbf{i}_{out})-\mathbf{u}.
\end{equation}

% \end{enumerate}

%%%%%%%%%%%%%%%%%%%%%%%%%%%%%%%%%%%%%%%%%%%%%%UCM
\subsection{Unified Camera Model}

Using the reformulation from \cite{DS}, the intrinsic parameters and distortion coefficient for the Unified Camera Model (UCM) are defined as follows:  $\mathbf{i}=(f_x,f_y,c_x,c_y,\alpha)$, where $\alpha \in [0,1]$.

The projection and unprojection functions for UCM are defined respectively as:

% \subsubsection{Projection function}
\begin{align}
\label{UCM_projection}
\mathbf{\pi}(\mathbf{x},\mathbf{i}) &= 
\begin{bmatrix}
f_x\frac{x}{\alpha d+(1-\alpha)z}\\ 
f_y\frac{y}{\alpha d+(1-\alpha)z}
\end{bmatrix}
+
\begin{bmatrix}
c_x\\ 
c_y
\end{bmatrix},
\\
\mathbf{\pi}^{-1}(\mathbf{u}, \mathbf{i}) &=
\frac{\xi+\sqrt{1+(1-\xi^2)r_u^2}}{1+r_u^2}
\begin{bmatrix}
   m_x \\
   m_y \\
   1
\end{bmatrix}-
\begin{bmatrix}
 0 \\
 0 \\
 \xi
\end{bmatrix},
\end{align}
where
\begin{align}
d &= \sqrt{x^2+y^2+z^2}, \\
r_u^2 &= m_x^2 + m_y^2, \\
\xi &= \frac{\alpha}{1-\alpha},\\
\begin{bmatrix}
   m_x \\
   m_y
\end{bmatrix}&=
\begin{bmatrix}
   \frac{u-c_x}{f_x}(1-\alpha)\\
   \frac{v-c_y}{f_y}(1-\alpha)
\end{bmatrix}.
\end{align}

The conditions for projection and unprojection are:

\begin{align}
\Omega &= \{\mathbf{x} \in \mathbb{R}^3 \mid z > -wd\},
\\
\Theta &= 
\begin{cases} 
\mathbb{R}^2 & \text{if } \alpha \leq 0.5, \\
\{\mathbf{u} \in \mathbb{R}^2 \mid r_u^2 \leq \frac{(1-\alpha)^2}{2\alpha-1}\} & \text{if } \alpha > 0.5.
\end{cases}
\\
w &= 
\begin{cases} 
\frac{\alpha}{1-\alpha}, & \text{if } \alpha \leq 0.5, \\
\frac{1-\alpha}{\alpha}, & \text{if } \alpha > 0.5,
\end{cases}
\end{align}

In the initialization step for UCM, $f_x, f_y, c_x$ and $c_y$ are inherited from existing parameters, and only the distortion coefficient $\alpha$ is initialized. Thus,
 $\chi = \alpha$, and $\mathbf{A}$ and $\mathbf{b}$ are derived from the projection function as follows:
 
% \subsubsection{Initialization}
\begin{align}
\mathbf{A} &=
\begin{bmatrix}
(d_1-z_1)\mathbf{u_1}-\mathbf{c}\\ 
\cdots \\
(d_n-z_n)\mathbf{u_n}-\mathbf{c}
\end{bmatrix},\\
\mathbf{b} &=
\begin{bmatrix}
\mathbf{f} \odot \mathbf{p}_1-z_1(\mathbf{u}_1-\mathbf{c})\\ 
\cdots\\ 
\mathbf{f} \odot \mathbf{p}_n-z_n(\mathbf{u}_n-\mathbf{c})
\end{bmatrix},
\end{align}
where $\mathbf{u}_i=[u_i, v_i]^T$,  $\mathbf{c}=[c_x, c_y]^T$,  $\mathbf{f}=[f_x, f_y]^T$,  $\mathbf{p}_i=[x_i, y_i]^T$, and $\odot$ is the Hadamard product.

The error term for optimization is modified for easier derivation of the Jacobian for the $\alpha$ term: 

% \subsubsection{Optimization}
\begin{equation}
\label{UCMcost}
\mathbf{e}(\mathbf{i})=\begin{bmatrix}
f_xx-(u-c_x)(\alpha d+(1-\alpha)z)\\ 
f_yy-(v-c_y)(\alpha d+(1-\alpha)z)
\end{bmatrix},
\end{equation}
and the Jacobians for $f_x, f_y, c_x, c_y$ and $\alpha$ can be easily obtained as:

\begin{align}
\frac{\partial{\mathbf{e}}}{\partial{[f_x,f_y]}}
&= \begin{bmatrix}
x & 0\\ 
0 & y
\end{bmatrix},\\
\frac{\partial{\mathbf{e}}}{\partial{[c_x,c_y]}}
&= \begin{bmatrix}
\alpha d+(1-\alpha)z & 0\\ 
0 & \alpha d+(1-\alpha)z
\end{bmatrix},\\
\frac{\partial{\mathbf{e}}}{\partial\alpha}
&=
\begin{bmatrix}
(z-d)(u-c_x)\\ 
(z-d)(v-c_y)
\end{bmatrix}.
\end{align}

%%%%%%%%%%%%%%%%%%%%%%%%%%%%%%%%%%%%%%%%%%%%EUCM
\subsection{Enhanced Unified Camera Model}

According to the redefinition of Enhanced Unified Camera Model (EUCM) from \cite{DS}, the parameters for the EUCM include an additional parameter $\beta$ to those of the UCM:
$\mathbf{i}=(f_x,f_y,c_x,c_y,\alpha, \beta)$ where $\alpha \in [0,1]$ , $\beta > 0$.

The projection and unprojection functions for EUCM are as follows:

% \subsubsection{Projection function}
% \begin{align*}
\begin{align}
\mathbf{\pi}(\mathbf{x},\mathbf{i}) &= 
\begin{bmatrix}
f_x\frac{x}{\alpha d+(1-\alpha)z}\\ 
f_y\frac{y}{\alpha d+(1-\alpha)z}
\end{bmatrix}
+
\begin{bmatrix}
c_x\\ 
c_y
\end{bmatrix},
\\
\mathbf{\pi}^{-1}(\mathbf{u}, \mathbf{i}) &=
\frac{1}{\sqrt{m_x^2+m_y^2+m_z^2}}
\begin{bmatrix}
   m_x \\
   m_y \\
   m_z
\end{bmatrix},
\end{align}
where
\begin{align}
r_u^2 &= m_x^2 + m_y^2,\\
\label{EUCMd}
d &= \sqrt{\beta(x^2+y^2)+z^2}, \\
\begin{bmatrix}
   m_x \\
   m_y \\
   m_z
\end{bmatrix}&=
\begin{bmatrix}
   \frac{u-c_x}{f_x}\\
   \frac{v-c_y}{f_y}\\
   \frac{1-\beta\alpha^2r_u^2}{\alpha\sqrt{1-(2\alpha-1)\beta r_u^2}+(1-\alpha)}
\end{bmatrix}.
\end{align}

% \end{align*}

The conditions for projection and unprojection are: 

\begin{align}
\Omega &= 
\begin{cases} 
\mathbb{R}^3 & \text{if } \alpha \leq 0.5, \\
\{\mathbf{x} \in \mathbb{R}^3 \mid z \geq \frac{(\alpha-1)(\alpha d+(1-\alpha)z)}{2\alpha-1}\} & \text{if } \alpha > 0.5,
\end{cases}
\\
\Theta &= 
\begin{cases} 
\mathbb{R}^2 & \text{if } \alpha \leq 0.5, \\
\{\mathbf{u} \in \mathbb{R}^2 \mid r^2 \leq \frac{1}{\beta(2\alpha-1)}\} & \text{if } \alpha > 0.5.
\end{cases}
\end{align}

In EUCM, $\beta$ is not linearly solved, so it is set to 1 during initialization, similar to UCM, to obtain the value of $\alpha$.

For optimization, EUCM can use the same cost function as UCM since $\beta$ only affects $d$ (\ref{EUCMd}). The derivatives $\frac{\partial{\mathbf{e}}}{\partial f_x}, \frac{\partial{\mathbf{e}}}{\partial f_y}, \frac{\partial{\mathbf{e}}}{\partial c_x}, \frac{\partial{\mathbf{e}}}{\partial c_y},$ and $\frac{\partial{\mathbf{e}}} {\partial \alpha}$ are the same as in UCM. Additionally, the derivative with respect to $\beta$ can be obtained as follows:

\begin{equation}
\frac{\partial{\mathbf{e}}}{\partial\beta}=
-\begin{bmatrix}
\alpha\frac{(x^2+y^2)(u-c_x)}{2\sqrt{\beta(x^2+y^2)+z^2}}\\ 
\alpha\frac{(x^2+y^2)(v-c_y)}{2\sqrt{\beta(x^2+y^2)+z^2}}
\end{bmatrix}.
\end{equation}

%%%%%%%%%%%%%%%%%%%%%%%%%%%%%%%%%%%%%%%%%%%%%%%%DS
\subsection{Double Sphere}

The parameters for the Double Sphere (DS) model include an additional parameter $\xi$ compared to the UCM:
$\mathbf{i}=(f_x,f_y,c_x,c_y, \alpha, \xi)$ where $\alpha \in [0,1]$.

The projection and unprojection functions for the DS model are as follows:

% \subsubsection{Projection function}
\begin{align}
\mathbf{\pi}(\mathbf{x},\mathbf{i}) &= 
\begin{bmatrix}
f_x\frac{x}{\alpha d_2+(1-\alpha)(\xi d_1+z)}\\ 
f_y\frac{y}{\alpha d_2+(1-\alpha)(\xi d_1+z)}
\end{bmatrix}
+
\begin{bmatrix}
c_x\\ 
c_y
\end{bmatrix},
\\
\mathbf{\pi}^{-1}(\mathbf{u}, \mathbf{i}) &=
\frac{m_z\xi+\sqrt{m_z^2+(1-\xi^2)r_u^2}}{m_z^2+r_u^2}
\begin{bmatrix}
   m_x \\
   m_y \\
   m_z
\end{bmatrix}-
\begin{bmatrix}
 0 \\
 0 \\
 \xi
\end{bmatrix},
\end{align}
where 
\begin{align}
d_1 &= \sqrt{x^2+y^2+z^2},\\
d_2 &= \sqrt{x^2+y^2+(\xi d_1+z)^2},\\
r_u^2 &= m_x^2 + m_y^2, \\
\begin{bmatrix}
   m_x \\
   m_y \\
   m_z
\end{bmatrix}&=
\begin{bmatrix}
   \frac{u - c_x}{f_x}\\
   \frac{v - c_y}{f_y}\\
   \frac{1-\alpha^2r_u^2}{\alpha\sqrt{1-(2\alpha-1)r_u^2}+1-\alpha}
\end{bmatrix}.
\end{align}

The conditions for each function are:

% \subsubsection{Conditions}
\begin{align}
\Omega &= \left\{ \mathbf{x} \in \mathbb{R}^3 \mid z > -w_2d_1 \right\},
\\
\Theta &= 
\begin{cases} 
\mathbb{R}^2 & \text{if } \alpha \leq 0.5, \\
\{\mathbf{u} \in \mathbb{R}^2 \mid r^2 \leq \frac{1}{2\alpha-1}\} & \text{if } \alpha > 0.5,
\end{cases}
\end{align}
where 
\begin{align}
w_2 &= \frac{w_1 + \xi}{\sqrt{2w_1\xi + \xi^2 + 1}},\\
w_1 &= 
\begin{cases} 
\frac{\alpha}{1-\alpha}, & \text{if } \alpha \leq 0.5, \\
\frac{1-\alpha}{\alpha}, & \text{if } \alpha > 0.5.
\end{cases}
\end{align}

Like EUCM, since $\xi$ cannot be solved linearly, it is set to 0, and initialization for $\alpha$  is conducted similarly to UCM.

For optimization, the cost function is modified as below:

% \subsubsection{Optimization}
\begin{align}
\mathbf{e}(\mathbf{i}) &= \begin{bmatrix}
f_xx-(u-c_x)(\alpha d_2+(1-\alpha)(\xi d_1+z))\\ 
f_yy-(v-c_y)(\alpha d_2+(1-\alpha)(\xi d_1+z))
\end{bmatrix},
\end{align}
and unlike EUCM, $\xi$ affects all terms except $f_x$ and $f_y$ thus the Jacobians for each term are recalculated as follows:

\begin{align}
\frac{\partial{\mathbf{e}}}{\partial{[f_x,f_y]}} &= \begin{bmatrix}
x & 0\\ 
0 & y
\end{bmatrix}, \\
\frac{\partial{\mathbf{e}}}{\partial{c_x}} &= \begin{bmatrix}
\alpha d_2+(1-\alpha)(\xi d_1+z) \\ 
0 
\end{bmatrix}, \\
\frac{\partial{\mathbf{e}}}{\partial{c_y}} &= \begin{bmatrix}
 0\\ 
 \alpha d_2+(1-\alpha)(\xi d_1+z)
\end{bmatrix}, \\
\frac{\partial{\mathbf{e}}}{\partial\alpha} &= \begin{bmatrix}
(\xi d_1+z-d_2)(u-c_x)\\ 
(\xi d_1+z-d_2)(v-c_y)
\end{bmatrix}, \\
\frac{\partial{\mathbf{e}}}{\partial\xi} &= -\begin{bmatrix}
(u-c_x)(\frac{\alpha d_1(\xi d_1+z)}{d_2}+(1-\alpha)d_1)\\ 
(v-c_y)(\frac{\alpha d_1(\xi d_1+z)}{d_2}+(1-\alpha)d_1)
\end{bmatrix}.
\end{align}

%%%%%%%%%%%%%%%%%%%%%%%%%%%%%%%%%%%%%%%%%%%%%%KB
\subsection{Kannala-Brandt Camera Model}

The intrinsic parameters and distortion coefficients for the Kannala-Brandt (KB) model are as follows:
$\mathbf{i}=(f_x,f_y,c_x,c_y,k_1,k_2,k_3,k_4)$.

The projection function for the KB model is defined as:

\begin{equation}
\label{KB_projection}
\mathbf{\pi}(\mathbf{x},\mathbf{i}) = 
\begin{bmatrix}
f_xd(\theta)\frac{x}{r}\\ 
f_yd(\theta)\frac{y}{r}
\end{bmatrix}
+
\begin{bmatrix}
c_x\\ 
c_y
\end{bmatrix},
\end{equation}
where
\begin{align}
r &= \sqrt{x^2+y^2},\\
\theta &= atan2(r,z), \\
d(\theta)&=\theta + k_1\theta^3+k_2\theta^5+k_3\theta^7+k_4\theta^9,
\end{align}
this projection is applicable in $\Omega = \mathbb{R}^3 \setminus \{[0,0,0]^T\}$.

The unprojection function for the KB model is:

\begin{equation}
\mathbf{\pi}^{-1}(\mathbf{u}, \mathbf{i}) = 
% \begin{bmatrix}
% \sin(\theta^*) & \sin(\theta^*) & \cos(\theta^*)
% \end{bmatrix} ^T
% \begin{bmatrix}
% \frac{m_x}{r_u} & \frac{m_y}{r_u} & 1
% \end{bmatrix}
\begin{bmatrix}
    \sin(\theta^*) \frac{m_x}{r_u} \\
    \sin(\theta^*) \frac{m_y}{r_u} \\
    \cos(\theta^*)
\end{bmatrix},
\end{equation}
where
\begin{align}
\label{KB_newton}
\theta^* &= d^{-1}(r_u),\\
r_u &= \sqrt{m_x^2 + m_y^2}, \\
\begin{bmatrix}
   m_x \\
   m_y
\end{bmatrix}&=
\begin{bmatrix}
   \frac{u-c_x}{f_x}\\
   \frac{v-c_y}{f_y}
\end{bmatrix}.
\end{align}

This unprojection is valid for all 2D space when  $d(\theta)$ is monotonic. The angle  $\theta^*$ satisfying Equation (\ref{KB_newton}) can be obtained using algorithms such as Newton-Raphson \cite{newton-raphson}.

For parameter initialization in optimization, $f_x, f_y, c_x,$ and $c_y$ are inherited from the input model, and only the remaining distortion coefficients are initialized. Thus, $\chi=[k_1,k_2,k_3,k_4]^T$, and $\mathbf{A}$ and $\mathbf{b}$ are derived from the projection function as follows:

\begin{align}
\mathbf{A}&=
\mathbf{1}_{\scriptsize 2n} \mathbf{v} ^T,\\
\mathbf{b}&=
\begin{bmatrix}
(\mathbf{u}_1 - \mathbf{c}) \odot (\mathbf{f} \odot \mathbf{p}_1)^{-1} r_1 - \theta_1 \mathbf{1}_{2}\\
% (u_1-c_x)\frac{r_1}{f_xx_1}-\theta_1\\ 
% (v_1-c_y)\frac{r_1}{f_yy_1}-\theta_1\\ 
\cdots\\ 
(\mathbf{u}_n - \mathbf{c}) \odot (\mathbf{f} \odot \mathbf{p}_n)^{-1} r_n - \theta_n \mathbf{1}_{2}
\end{bmatrix},
\end{align}
where $\mathbf{v} = [\theta^3, \theta^5, \theta^7, \theta^9] ^T$.

For optimization, $\mathbf{e(i)}$ can be defined as shown in Equation (\ref{generale}) and the Jacobian for each parameter is defined as follows:

\begin{align}
\frac{\partial{\mathbf{e}}}{\partial{[f_x,f_y,c_x,c_y]}}
&= \begin{bmatrix}
d(\theta)\frac{x}{r} & 0 & 1 & 0\\ 
0 & d(\theta)\frac{y}{r} & 0 & 1
\end{bmatrix},\\
\frac{\partial{\mathbf{e}}}{\partial{k_{1\cdots4}}}=\frac{\partial{\mathbf{e}}}{\partial{d(\theta)}}\frac{\partial{d(\theta)}}{\partial{k_{1\cdots4}}}&=\begin{bmatrix}
 f_x\frac{x}{r}\\ 
 f_y\frac{y}{r} 
\end{bmatrix}
\begin{bmatrix}
\theta^3 & \theta^5 & \theta^7 & \theta^9
\end{bmatrix}.
\end{align}

%%%%%%%%%%%%%%%%%%%%%%%%%%%%%%%%%%%%%%%%%%%%%%%%%%%OCC
\subsection{OCamCalib Camera Model}

The parameters for the OCamCalib (OCC) model are as follows:
$\mathbf{i}=(c,d,e,c_x,c_y,\mathbf{a}, \mathbf{k})$.

Unlike other models, the OCC model does not use focal lengths $f_x$ and $ f_y$ but instead employs an affine transformation matrix $[c, d; e, 1]$ for sensor alignment.  $\mathbf{a}=(a_0, a_1, a_2, a_3, a_4)$ are the coefficients of the polynomial function used in the unprojection function, and $\mathbf{k}=(k_0, k_1,k_2,k_3,k_4, \cdots, k_p)$ are the coefficients for the polynomial function used in the projection function.

The projection and unprojection functions for the OCC model are defined as follows:

% \subsubsection{Projection function}
\begin{align}
\label{OCC_projection}
\mathbf{\pi}(\mathbf{x},\mathbf{i}) &= 
\begin{bmatrix}
c&d\\ 
e&1
\end{bmatrix}
\begin{bmatrix}
d(\theta)\frac{x}{r}\\ 
d(\theta)\frac{y}{r}
\end{bmatrix}
+
\begin{bmatrix}
c_x\\ 
c_y
\end{bmatrix},
\\
\mathbf{\pi}^{-1}(\mathbf{u}, \mathbf{i}) &=
\frac{1}{\sqrt{m_x^2+m_y^2+m_z^2}}
\begin{bmatrix}
   m_x \\
   m_y \\
   m_z
\end{bmatrix},
\end{align}
where
% \subsubsection{Un-projection function}
\begin{align}
r &= \sqrt{x^2+y^2},\\
r_u &= \sqrt{m_x^2 + m_y^2},\\
\theta &= atan(\frac{z}{r}), \\
d(\theta) &= k_0+k_1\theta+k_2\theta^2+k_3\theta^3 \cdots k_p\theta^p, \\
\begin{bmatrix}
   m_x \\
   m_y
\end{bmatrix}&=
\begin{bmatrix}
   c &d\\
   e&1
\end{bmatrix}^{-1}
\begin{bmatrix}
   u-c_x\\
   v-c_y
\end{bmatrix}
,\\
m_z&=a_0+a_1r_u+a_2r_u^2+a_3r_u^3+a_4r_u^4.
\end{align}

% \subsubsection{Initialization}
As in the original paper \cite{OCC}, initialization sets $c=1, d=0$ and $e=0$ focusing on  $\mathbf{a}$ because the order of the polynomial function used in the unprojection is experimentally determined to be 4, while the order of the polynomial function for projection is not specified. Therefore, $\mathbf{A}$ and $\mathbf{b}$ are also derived differently, using the unprojection function:

\begin{align}
\chi&= [a_0,a_1,a_2,a_3,a_4]^T,\\
\mathbf{A}&=
\mathbf{1}_{\scriptsize 2n} \mathbf{o} ^T,\\
\mathbf{b}&=
\begin{bmatrix}
z_1 (\mathbf{u}_1 - \mathbf{c}) \odot \mathbf{p}_1^{-1}\\ 
\cdots\\ 
z_n (\mathbf{u}_n - \mathbf{c}) \odot \mathbf{p}_n^{-1}
\end{bmatrix},
\end{align}
where $\mathbf{o}=[1, r_u, r_u^2, r_u^3, r_u^4]^T$.

The error term for optimization, derived from the unprojection function, is:

\begin{equation}
\mathbf{e}(\mathbf{i})=\begin{bmatrix}
(u-c_x)-d(v-c_y)-m_z(c-de)x/z\\ 
e(u-c_x)+c(v-c_y)-m_z(c-de)y/z
\end{bmatrix}.
\end{equation}

Given that $m_z$ is a function of $(c, d, e)$ and goes up to the 8th degree, linear approximation for the Jacobian is challenging. Therefore, using $c=1, d=0$ and $e=0$  the cost function is redefined as:

\begin{equation}
\mathbf{e}(\mathbf{i})=\begin{bmatrix}
(u-c_x)-\tilde{m}_z x/z\\ 
(v-c_y)-\tilde{m}_z y/z
\end{bmatrix},
\end{equation}
where
\begin{align}
\tilde{r}_u &= \sqrt{\tilde{m}_x^2 + \tilde{m}_y^2},\\
\tilde{m}_z&=a_0+a_1\tilde{r}_u+a_2\tilde{r}_u^2+a_3\tilde{r}_u^3+a_4\tilde{r}_u^4,
\\
\begin{bmatrix}
   \tilde{m}_x \\
   \tilde{m}_y
\end{bmatrix}&=
\begin{bmatrix}
   1 &0\\
   0&1
\end{bmatrix}
\begin{bmatrix}
   u-c_x\\
   v-c_y
\end{bmatrix}.
\end{align}

Since the parameters $(c,d,e)$ are not directly optimized, their role in correcting sensor misalignment through affine transformation is expected to be compensated by the parameters $c_x$ and $c_y$ Thus, the final Jacobian for OCC's parameters is defined accordingly,

% \begin{equation}
% \frac{\partial{\mathbf{e}}}{\partial{c}}
% = \begin{bmatrix}
% -m_z m_x\\ 
% -m_z m_y + (v-c_y)
% \end{bmatrix},
% \end{equation}
% \begin{equation}
% \frac{\partial{\mathbf{e}}}{\partial{d}}
% = \begin{bmatrix}
% -(v-c_y)+e m_z m_x\\ 
% e m_z m_y
% \end{bmatrix},
% \end{equation}
% %%%%%%%%%%%%%%%%%%%%%%%%%%%%%%%%%%%%% partial e/ partial e
% \begin{equation}
% \frac{\partial{\mathbf{e}}}{\partial{e}}
% = \begin{bmatrix}
% d m_z m_x\\ 
% -m_z m_y - (u-c_x)
% \end{bmatrix},
% \end{equation}
%%%%%%%%%%%%%%%%%%%%%%%%%%%%%%%%%%%%% partial e/ partial cx, cy
\begin{align}
\frac{\partial{\mathbf{e}}}{\partial{[c_x,c_y]}}
&= -\begin{bmatrix}
1 & 0\\ 
0 & 1
\end{bmatrix},
\\
%%%%%%%%%%%%%%%%%%%%%%%%%%%%%%%%%%%%% partial e/ partial a_0...4
\frac{\partial{\mathbf{e}}}{\partial a_{0 \cdots 4}}
&=
-\begin{bmatrix}
\frac{x}{z}\\ 
\frac{y}{z}
\end{bmatrix}
\begin{bmatrix}
1&\tilde{r}_u&\tilde{r}_u^2&\tilde{r}_u^3&\tilde{r}_u^4
\end{bmatrix}.
\end{align}

Subsequently, as proposed in the paper \cite{OCC}, $p$ is automatically estimated by solving the following linear equation to minimize the reprojection error,

\begin{equation}
\begin{bmatrix}
\frac{\mathbf{p}_1}{r_1} \\
\frac{\mathbf{p}_n}{r_2} \\
\vdots \\
\frac{\mathbf{p}_n}{r_n}
\end{bmatrix}
\begin{bmatrix}
1 \\ \theta \\ \vdots \\ \theta^p 
\end{bmatrix}^T
% \frac{x_1}{r_1} & \frac{x_1}{r_1}\theta & \frac{x_1}{r_1}\theta^2 & \cdots & \frac{x_1}{r_1}\theta^p\\ 
% \frac{y_1}{r_1} & \frac{y_1}{r_1}\theta & \frac{y_1}{r_1}\theta^2 & \cdots & \frac{y_1}{r_1}\theta^p\\ 
% \vdots & \vdots & \vdots & \ddots & \vdots\\
% \frac{x_n}{r_n} & \frac{x_n}{r_n}\theta & \frac{x_n}{r_n}\theta^2 & \cdots & \frac{x_n}{r_n}\theta^p\\ 
% \frac{y_n}{r_n} & \frac{y_n}{r_n}\theta & \frac{y_n}{r_n}\theta^2 & \cdots & \frac{y_n}{r_n}\theta^p
% \end{bmatrix}
\begin{bmatrix}
k_0\\ 
k_1\\ 
\vdots\\
k_p
\end{bmatrix}
=
\begin{bmatrix}
\mathbf{u}_1-\mathbf{c}\\
\mathbf{u}_2-\mathbf{c}\\
\vdots\\
\mathbf{u}_n-\mathbf{c}
\end{bmatrix}.
\end{equation}

%%%%%%%%%%%%%%%%%%%%%%%%%%%%%%%%%%%%%%%%%%%%%%RadTan
\subsection{Radial-Tangential Distortion Model}

The intrinsic parameters and distortion coefficients for the Radial-Tangential (RT) distortion model are as follows:
$\mathbf{i}=(f_x,f_y,c_x,c_y,k_1,k_2,k_3,p_1, p_2)$.

The projection function for the RT model is:

\begin{equation}
\mathbf{\pi}(\mathbf{x},\mathbf{i}) = 
\begin{bmatrix}
f_x x''\\ 
f_y y''
\end{bmatrix}
+
\begin{bmatrix}
c_x\\ 
c_y
\end{bmatrix},
\end{equation}
where
\begin{align}
r^2&=x'^2+y'^2,\\
r' &= 1+k_1r^2+k_2r^4+k_3r^6,\\
\label{RadTan_newton}
\begin{bmatrix}
   x'' \\
   y''
\end{bmatrix}&=
\begin{bmatrix}
   r'x'+2p_1x'y'+p_2(r^2+2x'^2)\\
   r'y'+2p_2x'y'+p_1(r^2+2y'^2)
\end{bmatrix},
\\
\begin{bmatrix}
   x' \\
   y'
\end{bmatrix}&=
\begin{bmatrix}
   \frac{x}{z}\\
   \frac{y}{z}
\end{bmatrix}.
\end{align}

The unprojection function $\mathbf{\pi}^{-1}(\mathbf{u}, \mathbf{i})$ for the RT model recovers $x'$ and $y'$ to satisfy given $x''$ and $y''$ . This restoration process is non-linear and can be computed using methods such as the Newton-Raphson. The Jacobian for $x'$ and $y'$ that satisfy Equation (\ref{RadTan_newton}) is as follows:

\begin{equation}
\mathbf{J}_{\cdot 1} = \begin{bmatrix}
    r' + 2x'(k_1+2r^2k_2+3r^4k_3)+2p_1y'+6p_2x' \\
    2x'y'(k_1+2r^2k_2+3r^4k_3)+2p_1x'+2p_2y'
\end{bmatrix},
\end{equation}
\begin{equation}
\mathbf{J}_{\cdot 2} = \begin{bmatrix}
    2x'y'(k_1+2r^2k_2+3r^4k_3)+2p_2y'+2p_1x' \\
    r' + 2y'(k_1+2r^2k_2+3r^4k_3)+2p_2x'+6p_1y'
\end{bmatrix}.
\end{equation}

For optimization, parameter initialization is conducted as follows: $f_x, f_y, c_x$ and $c_y$ are inherited from the input model, while the remaining distortion coefficients are initialized. To simplify the model, $p_1$ and $p_2$ are initialized to zero, and the focus is on $\chi=[k_1,k_2,k_3]^T$. $\mathbf{A}$ and $\mathbf{b}$ are derived from the projection function as follows:

\begin{align}
\mathbf{A}&=
\mathbf{1}_{\scriptsize 2n} \mathbf{r} ^T,
\\
\mathbf{b}&=
\begin{bmatrix}
(\mathbf{u}_1 - \mathbf{c}) \odot (\mathbf{f} \odot \mathbf{p}'_1)^{-1} - \mathbf{1}_{2}\\
% (u_1-c_x)\frac{r_1}{f_xx_1}-\theta_1\\ 
% (v_1-c_y)\frac{r_1}{f_yy_1}-\theta_1\\ 
\cdots\\ 
(\mathbf{u}_n - \mathbf{c}) \odot (\mathbf{f} \odot \mathbf{p}'_n)^{-1} -  \mathbf{1}_{2}
\end{bmatrix},
\end{align}
where $\mathbf{r} = [r^2, r^4, r^6]^T$, and $\mathbf{p}'_i=[x', y']^T$.

The error term for optimization, $\mathbf{e(i)}$ can be defined as in Equation (\ref{generale}) with the Jacobian for each parameter defined as:

\begin{align}
\frac{\partial{\mathbf{e}}}{\partial{[f_x,f_y,c_x,c_y]}}
&= \begin{bmatrix}
x'' & 0 & 1 & 0\\ 
0 & y'' & 0 & 1
\end{bmatrix},
\\
\frac{\partial{\mathbf{e}}}{\partial{[k_1, k_2, k_3]}}
&= \begin{bmatrix}
x'r^2 & x'r^4 & x'r^6\\ 
y'r^2 & y'r^4 & y'r^6
\end{bmatrix},
\\
\frac{\partial{\mathbf{e}}}{\partial{[p_1, p_2]}}
&= \begin{bmatrix}
2x'y' & r^2+x'^2\\ 
r^2+2y'^2 & 2x'y'
\end{bmatrix}.
\end{align}

%%%%%%%%%%%%%%%%%%%%%%%%%%%%%%%%%%%%%%%%%%%%%%%%%%%Custom Model
\subsection{Custom Camera Model}

The proposed FCA module is modularized to support unprojection, projection, initialization, and optimization for various camera models. For example, the fisheye camera model provided by the WoodScape dataset \cite{woodscape} is similar to KB model but with slight differences.

The implementation of $d(\theta)$ for the WoodScape dataset's model is defined as follows:

\begin{equation}
d(\theta) = k_1\theta + k_2\theta^2 + k_3 \theta^3 + k_4 \theta^4.
\end{equation}

Apart from this, the projection, unprojection, initialization, and optimization processes can be conducted in the same manner as with the KB model.

% Experiment
\section{Experiment}
\label{sec:experiment}

%\Implementation Details
This system was implemented in C++ on a system equipped with an AMD Ryzen 7 5800U CPU and 16GB RAM. Optimization was performed using the Ceres Solver\footnote{http://ceres-solver.org/}.

\subsection{Evaluation}

We first acquired the sample point $N$ for our proposed model.
To obtain the sample points, we first compared parameter error and execution time. The samples were extracted by uniformly dividing the given image size into grid cells to ensure $N$ samples were evenly distributed.
Parameter error was calculated using the L2-norm of $\mathbf{i}^* - \hat{\mathbf{i}}$.

\noindent\textbf{Experiment on Kalibr Dataset}
The experiments utilized the Kalibr dataset \cite{kalibr}, and since the Kalibr calibration toolkit allows direct calibration for the KB, EUCM, DS, and RT models, we used parameters obtained with the Kalibr toolkit as the ground truth $\mathbf{i}^*$ Here, $\hat{\mathbf{i}}$ refers to the output model converted from the input model, with input and output model pairs linked by a hyphen (i.e., EUCM-DS).

\begin{figure}[!tb]
      \centering
      \includegraphics[width=3in]{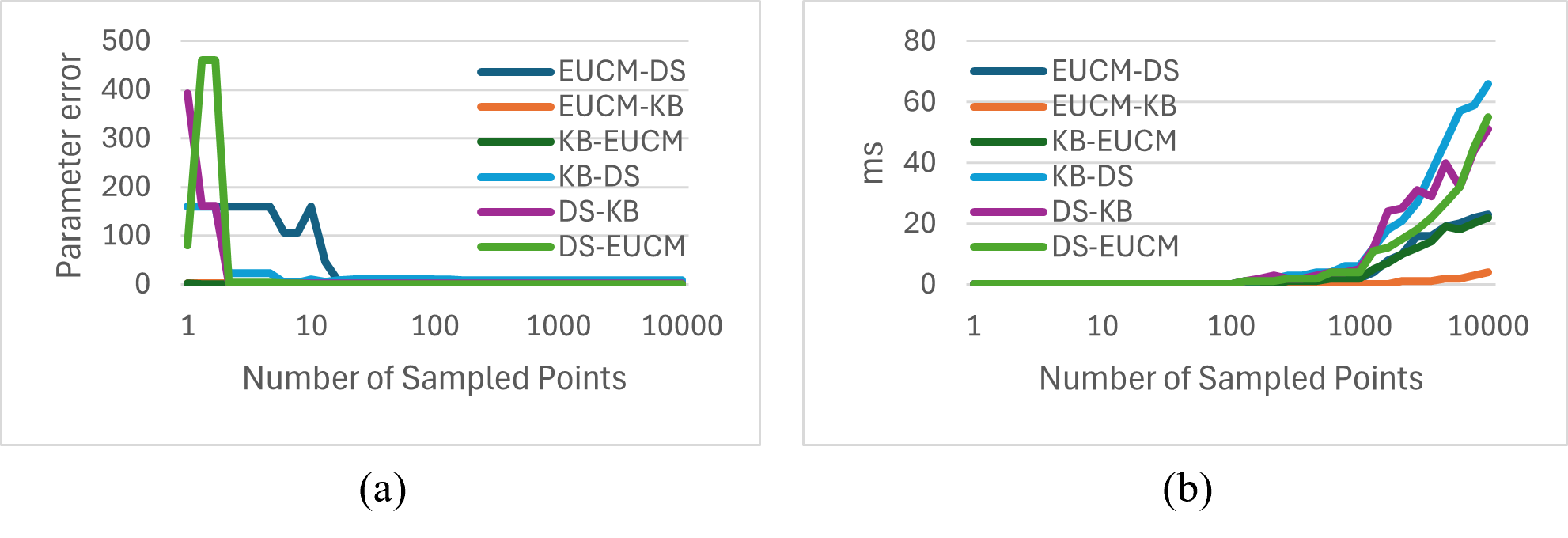}

      \caption{For the comparison of parameter error and execution time per sample point $N$ , the results are presented in two parts: (a) parameter error per sample point and (b) execution time per sample point. The notation A-B indicates the conversion from input model A to output model B. The parameter error begins to saturate at approximately $N=30$, and the execution time starts to significantly increase from around $N=1000$.}
      \label{figure2}
   \end{figure}

Figure \ref{figure2} shows the experimental results, indicating that parameter error saturates around $N=30$ The speed remains within 10 ms up to $N=1000$ and then increases linearly.

\begin{figure*}[!tb]
      \centering
      \includegraphics[width=\textwidth]{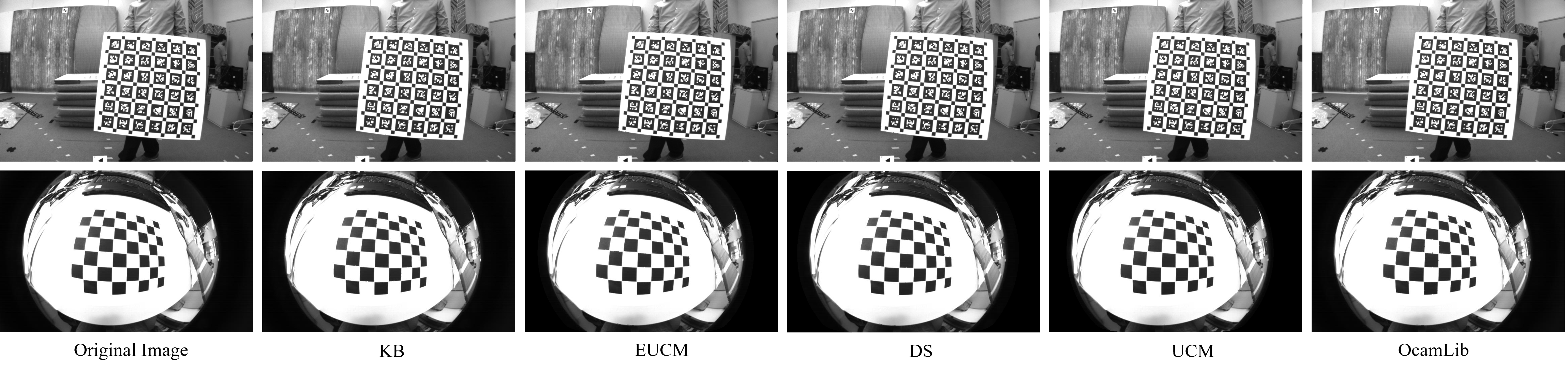}

      \caption{An example of recovered images. The first row is from the Kalibr dataset, and the second row is from the OCamCalib dataset. The input model is fixed as KB, and the output model corresponds to each column of the image.}
      \label{figure3}
   \end{figure*}

We observed more detailed results at $N=500$. 
The metrics for the ground truth output model relative to the input-output model include PSNR, SSIM \cite{PSNR_SSIM}, Reprojection Error (RE), and Parameter Error (PE) calculated using the original image and the image recovered using the output model.
The recovered image was obtained by unprojecting all pixels of the original image using the input model and then projecting them using the output model. Figure \ref{figure3} shows an example of a recovered image.

The experimental results are summarized in Table \ref{table1}. The conversion results of our proposed model show that RE is close to zero, and the PSNR and SSIM with the original image are excellent, along with fast estimation within 4 ms and a Parameter Error averaging 2.63, similar to the results obtained with the actual calibration toolkit. The results show that the conversions between the EUCM and KB models, as well as the DS model, are satisfactory.

Particularly for the RT model, which is a distortion modeling of the pinhole model, there is a performance degradation in the conversion from fisheye camera models compared to other models. Especially in the conversion to the DS model, there tends to be a significant discrepancy in the estimation of the focal length from the actual value, resulting in a higher final parameter error compared to other conversion models.

\noindent\textbf{Experiment on OcamCalib Dataset}
For experiments with the OCC model, we utilized 190-degree large FOV images from the OCamCalib dataset\footnote{https://sites.google.com/site/scarabotix/ocamcalib-omnidirectional-camera-calibration-toolbox-for-matlab}. Ground truth for the OCC was obtained using the OCamCalib calibration toolkit. However, ground truth for other models such as KB, EUCM, and DS could not be obtained as the toolkit does not support these models. Therefore, we acquired the input model as $\textbf{i}_{in} = \text{FCA}(\textbf{i}^*_{OCC})$ and used this input model to again obtain $\hat{\textbf{i}}_{OCC}$. In this experiment, the RT model was not compared due to its use of a large angle FoV.

The experimental results for OCC are shown in Table \ref{table2}. The results showed that the parameters estimated for various input models yielded an average PSNR of 34.5, SSIM of 0.84, RE of 0.33, and PE of 4.76, which are respectable outcomes.

\begin{table}[!tb]
\caption{Result on Kalibr dataset(N=500). Note that RE is the Reprojetion Error and PE is the Parameter Error.}
\label{table1}
\begin{center}
\begin{tabular}{lccccc}
\hline
\multicolumn{1}{c}{In-out model} & Time(ms) & PSNR & SSIM & RE & PE \\
\hline
                    EUCM-DS      &   2        & 38.2773         & 0.9975        &  7.75e-06 &  4.0964  \\
                    EUCM-KB      & \textbf{1} & 36.5411           & 0.9965        &  \textbf{6.87e-10} &  1.0779  \\
                    EUCM-RT  &   5        & 31.5081           & 0.9873        &  4.63e-05 &  2.5740  \\
                    KB-EUCM      & \textbf{1} & 40.4105         & 0.9975        &  0.02354  &  \textbf{0.5961}  \\
                    KB-DS        &   3        & 40.082         & 0.9974        &  0.02275  &  8.3069 \\
                    KB-RT    &   6        & 40.943         & 0.9978        &  0.2617  &  3.5305 \\
                    DS-KB        &   3        & 39.8845         & 0.9968        &  1.87e-05 &  1.4905  \\
                    DS-EUCM      &   2        & \textbf{42.4375}& \textbf{0.9981} &  0.0024   &  0.6312  \\
                    DS-RT      &   4        & 28.8887 & 0.9584 &  15.1505   &  157.024  \\
                    
                    RT-EUCM      &   4        & 40.5359 & 0.9968 &  0.7922   &  8.1594  \\
                    RT-KB      &   2        & 41.6988& 0.9979 &  0.1031   &  2.1977  \\
                    RT-DS      &   2        & 38.4396& 0.9961 &  0.9697   &  195.222  \\
                                 \hline
                                 
\end{tabular}
\end{center}
\end{table}

\begin{table}[!tb]
\caption{Result on OCamCalib dataset(N=500). RE indicates the Reprojetion Error and PE represents the Parameter Error.}
\label{table2}
\begin{center}
\begin{tabular}{lccccc}
\hline
\multicolumn{1}{c}{In-out model} & Time(ms) & PSNR & SSIM & RE & PE \\
\hline
KB-OCC     &  4          & \textbf{38.3962}  & 0.8227           & 0.1777 &2.181              \\
UCM-OCC    &  5         & 31.9255            & 0.8035           & 0.9480 & 13.7598             \\
EUCM-OCC   &  \textbf{3}  & 33.0033          & \textbf{0.9244}  & \textbf{0.0468} & \textbf{0.3695}   \\
DS-OCC     &  4         &  34.7245           & 0.8246           & 0.1546 & 2.7296 \\
                                 \hline
                                 
\end{tabular}
\end{center}
\end{table}

\noindent \textbf{Comparison with the State-of-the-Art Method}
We compared the accuracy of the fisheye camera model conversion method available in libPeR \cite{libPer} with our proposed method. For a pairwise comparison, parameters estimated up to the UCM model in the libPeR paper were used to estimate values for the OCC model and were compared with those from libPeR.

The ground truth for the given OCC model is $a^*_0 = 131.0074, a^*_1=0, a^*_2=-0.0018, c^*_x = 516.4379, c^*_y = 383.014$ , and the order of polynomial is set to 2, obtainable from the OCamCalib calibration tool. For details, refer to the relevant paper \cite{libPer}.

The values obtained for the UCM model in libPeR are $\hat{\gamma}_x=259.889, \hat{\gamma}_y=259.335, \hat{c}_x=514.168, \hat{c}_y=382.797, \hat{\xi}=0.975$; through model reformulation $\xi=\frac{\alpha}{1-\alpha}, \gamma_x = \frac{f_x}{1-\alpha}, \gamma_y = \frac{f_y}{1-\alpha}$, please refer to \cite{DS}, Equation (7)), we can derive $\tilde{f}_x=131.5893, \tilde{f}_y=131.3089, \tilde{c}_x=514.168, \tilde{c}_y=382.797, \tilde{\alpha} = 0.4937$.

Using these UCM parameter values, libPeR obtained $\hat{a}_0 = 131.46, \hat{a}_1=0, \hat{a}_2=-0.0018$, while our model yielded $\tilde{a}_0 = 130.809, \tilde{a}_1=0.01238, \tilde{a}_2=0.00186$. 

The comparison of the distortion coefficients with the ground truth is shown in Table \ref{table3}. The experimental results indicate that our proposed method achieves higher accuracy in terms of RMSE.

\begin{table}[b]
\caption{Comparison with the SOTA method on OCamCalib dataset}
\label{table3}
\begin{center}
\begin{tabular}{lcccc}
\hline
\multicolumn{1}{c}{Method} & $\Delta a_0$ & $\Delta a_1$ & $\Delta a_2$ & RMSE \\
\hline
libPeR \cite{libPer} &  0.4526 & \textbf{0}  & \textbf{0}           & 0.2613              \\
Ours    &  \textbf{0.1984}  & 0.0124  & 0.0037    & \textbf{0.1148}             \\
                                 \hline
                                 
\end{tabular}
\end{center}
\end{table}

\subsection{Application}

We validated the performance of our proposed model for converting fisheye camera models in actual applications.

Fisheye ORB-SLAM \cite{FisheyeORBSLAM} has adopted the EUCM model for fisheye cameras. Utilizing the KB model provided by the TUM Visual Inertial(VI) Dataset \cite{tum-vi-dataset}, we performed fisheye ORB-SLAM using both the KB-EUCM model parameters derived from converting the KB model and the directly acquired EUCM model parameters from the $calib-cam1$ sequence of the VI Dataset. We calculated the Absolute Pose Error (APE) and Relative Pose Error (RPE) for the acquired keyFrame trajectories using the evo package \cite{evo}. For the experiments, sequences $corridor4$ and $room2$ from the VI Dataset were used, as these relatively small spaces with potential for loop closing are less likely to exhibit significant trajectory divergence due to the randomness introduced by RANSAC in fisheye ORB-SLAM.

The experimental results, as shown in Table \ref{table4}, indicate that the difference between the directly acquired EUCM model and our KB-EUCM model is approximately 3 cm in average APE and about 0.1 cm in RPE. These differences are within the error margins typically expected in the system, confirming that our model’s conversion of fisheye camera models is performed correctly.

\begin{figure}[!tb]
      \centering
      \includegraphics[width=3in]{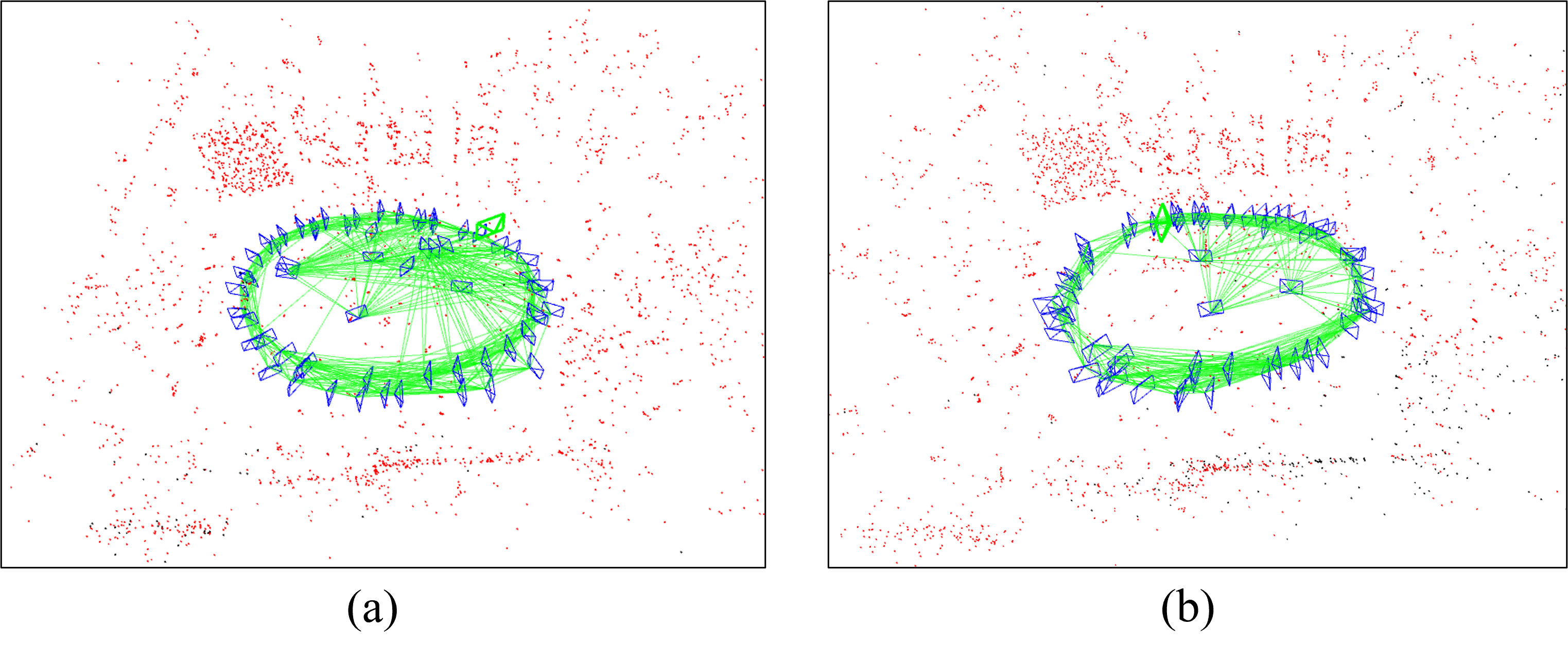}

      \caption{An example of fisheye ORB SLAM. (a) Results obtained using directly acquired EUCM parameters. (b) Results obtained using EUCM parameters acquired through KB-EUCM conversion performed by our model.}
      \label{figure4}
   \end{figure}

Figure \ref{figure4} shows the results of fisheye ORB SLAM performed with directly acquired EUCM parameters and with EUCM parameters obtained through KB-EUCM conversion. The result images demonstrate that normal SLAM operations are successfully carried out with the converted EUCM parameters, confirming that the trajectory and structure are well generated.

\begin{table}[!tb]
\caption{Comparison of trajectories obtained through fisheye ORB SLAM with EUCM and KB-EUCM model}
\label{table4}
\begin{center}
\begin{tabular}{lcc}
\hline
\multicolumn{1}{c}{Sequence} & APE (rmse/std) & RPE (rmse/std) \\
\hline
corridor4                    & 0.0293/0.0167 m  & 0.0097/0.008 m   \\
room2                        & 0.0013/0.0004 m  & 0.0021/0.001 m  \\
\hline
\end{tabular}
\end{center}
\end{table}

% See \cite{ref1,ref2,ref3,ref4,ref5} for resources on formatting math into text and additional help in working with \LaTeX .

\subsection{Limitation}
Our proposed FCA module acquires parameters for the output model from the input model. This process utilizes environmental information recovered from the input model. Consequently, the quality of the calibrated parameters of the input model naturally affects the results of the output model. Therefore, the quality of the output model improves as the accuracy of the input model's calibration results increases.

\section*{Acknowledgement}
This research was supported by StradVision. We appreciate all the supports of StradVision members who provided insight and expertise. The contents are solely the responsibility of the authors.

\section{Conclusion}
\label{sec:conclude}

We have proposed the Fisheye-Calib-Adapter, a tool designed to facilitate the easy conversion of fisheye camera models. Our system can quickly and accurately estimate parameters for the output model based solely on the intrinsic parameters of the camera model to be converted, without the need for any image set. Our method supports widely used models such as UCM, EUCM, Double Sphere, Kannala-Brandt, OCamCalib, and Radial-Tangential, and provides an interface for other custom models. The converted model parameters obtained using our system can be directly applied to applications like SLAM. We believe that our module will enable researchers to bridge the gap in fisheye camera models and be used in a variety of studies, as it allows the acquisition of parameters for the desired model without the need for recalibration.

%%%%%%%%%%%%%%%%%%%%%%%%%%%%%%%%%%%%%%%%%%%%%%%%%%%%%%%%%%%%%%%%%%%%%%%%%%%%%%%%

\bibliographystyle{IEEEtranBST/IEEEtran}
\bibliography{main}

\end{document}